\documentclass[fleqn,a4paper,12pt]{article}
\usepackage{amssymb}
\usepackage{amsmath}
\usepackage[dvips]{graphicx}
\usepackage{lscape}
\usepackage{a4}
\usepackage{epsfig}

\numberwithin{equation}{section}

\newcommand{\beq}{\begin{equation}} \newcommand{\eeq}{\end{equation}}
\newcommand{\bea}{\begin{array}} \newcommand{\eea}{\end{array}}
\newcommand{\ri}{{\mathrm i}}

\catcode`@=11 \long
\def\@caption#1[#2]#3{\par\addcontentsline{\csname
ext@#1\endcsname}{#1} {\protect\numberline{\csname
the#1\endcsname}{\ignorespaces #2}} \begingroup \small
\@parboxrestore \@makecaption{\csname fnum@#1\endcsname}
{\ignorespaces #3}\par \endgroup} \catcode`@=12

\newcommand{\p}{\partial}

\begin{document}
\allowdisplaybreaks
\begin{titlepage} \vskip 2cm

\begin{center} {\Large\bf Exact solvability of PDM systems with extended Lie symmetries}
\footnote{E-mail: {\tt nikitin@imath.kiev.ua} }\vskip 3cm {\bf {A.
G. Nikitin } \vskip 5pt {\sl Institute of Mathematics, National
Academy of Sciences of Ukraine,\\ 3 Tereshchenkivs'ka Street,
Kyiv-4, Ukraine, 01004\\}}\end{center} \vskip .5cm \rm

\begin{abstract}{It is shown that all PDM Schr\"odinger equations
admitting more than five dimensional Lie symmetry algebras (whose completed list can be found in paper~[{\it J.~Math. Phys.} {\bf 58}, , 083508 (2017)]
 are
exactly solvable. The corresponding exact solutions are presented. The supersymmetric aspects of the exactly solvable systems are
discussed.}\end{abstract}
\end{titlepage}

\section{  Introduction.} 
Group classification of differential equations consists in the
specification of non-equivalent classes of such equations which
possess the same symmetry groups. It is a rather attractive research
field which has both fundamental and application values.

A perfect example of group classification of fundamental equations
of mathematical physics was presented by Boyer~\cite{Boy} who had
specified all inequivalent Schr\"odinger equations with time
independent potentials admitting symmetries with respect to Lie
groups, see also papers \cite{Hag,Nied,And} where
particular important symmetries were discussed, and paper \cite{NB} were the Boyer results are corrected. These old results
have a big impact since include a priori information about all
symmetry groups which can be admitted by the fundamental equation of
quantum mechanics. Let us mention also that the nonlinear
Schr\"odinger equation as well as the generalized Ginsburg--Landau
quasilinear equations have been classified also \cite{pop,N1} as well as symmetries of more general systems of reaction-diffusion equations \cite{NW1,NW2}. For general discussion of {\it supersymmetries} of Schr\"odinger equation see, e.g., papers \cite{bek1,bek2,AK,AN1}.

In contrary, the group classification of Schr\"odinger equations
with position dependent mass (PDM) was waited for a very long time.
There were many papers devoted to PDM Schr\"odinger equations with
particular symmetries, see, e.g., \cite{11,rac,Koch,Cru}. But the complete group classification of
these equations appears only recently in papers \cite{NZ}  and \cite{NZ2,NN} for the stationary and time dependent
equations correspondingly. A systematic search for the higher order symmetries if the PDM systems started in paper \cite{N2}. So late making of such important job
have to cause the blame for experts in group analysis of
differential equations, taking into account the fundamental role
played by such equations in modern theoretical physics!

Let us remind that the PDM Schr\"odinger equations are requested for
description of various condensed-matter systems such as
semiconductors, quantum liquids, and metal clusters, quantum wells,
wires and dots, super-lattice band structures, etc., etc.

It happens that the number of PDM systems with different Lie
sym-metries is rather extended. Namely, in~\cite{NN} seventy classes
of such systems are specified. Twenty of them are defined up to
arbitrary parameters, the remaining fifty systems include arbitrary
functions.

The knowledge of all Lie groups which can be admitted by the PDM
Schr\"odinger equations has both fundamental and application values.
In particular, when construct the models with a priory requested
symmetries we can use the complete lists of inequivalent PDM systems
presented in~\cite{NZ2} for $d=2$ and~\cite{NN} for $d=3$. Moreover, in
many cases a~sufficiently extended symmetry induces integrability or
exact solvability of the system, and just this aspect will be
discussed in the present paper.

It will be shown that all PDM systems admitting six parametric Lie
groups of symmetries or more extended symmetries are exactly
solvable. Moreover, the complete sets of solutions of the
corresponding stationary PDM Schr\"odinger equations will be
presented explicitly.

There exist a tight connection between the complete solvability and
various types of higher symmetries and supersymmetries. We will see
that extended Lie symmetries also can cause the exact solvability.
More-over, the systems admitting extended Lie symmetries in many
cases are supersymmetric and superintegrable.

\section{ PDM Schr\"odinger equations with extended Lie sym\-met\-ries.}
In paper \cite{NN} we present the group classification of PDM
Schr\"odinger equations
\begin{gather}\label{seq}L\psi\equiv\left(\ri\frac{\p}{\p
t}-H\right)\psi=0,\end{gather} where $H$ is the PDM Hamiltonian of
the following generic form
\begin{gather}\label{B1} H=\tfrac14\big(m^\alpha p_a m^{\beta}p_am^\gamma+
m^\gamma p_a m^{\beta}p_am^\alpha\big)+ \hat V,\quad
p_a=-\ri\frac{\p}{\p x_a}.\end{gather} Here $m=m({\bf x})$ and $\hat
V=\hat V({\bf x})$ are the mass and potential depending on spatial
variables ${\bf x}=(x_1,x_2,x_3)$, and summation w.r.t.\ the
repeating indices $a$ is imposed over the values $a=1,2,3$. In
addition, $\alpha$, $\beta$ and $\gamma$ are the ambiguity parameters
satisfying the condition $\alpha+\beta+\gamma=-1$.

The choice of values of the ambiguity parameters can be motivated by
physical reasons, see a short discussion of this point in~\cite{NN}.

Hamiltonian (\ref{B1}) can be rewritten in the following more compact form
\begin{gather}\label{B2}H= \tfrac12p_a fp_a+ V,\end{gather}
where
\begin{gather}\label{B3} V=\hat V+\tfrac14(\alpha+\gamma)f_{aa}+
\alpha\gamma\frac{f_a f_a}{2f}\end{gather} with $f=\frac1m,\
f_a=\frac{\p f}{\p x_a}$ and $f_{aa}=\Delta f=\frac{\p f_a}{\p
x_a}$.

In the following text just representation (\ref{B3}) will be used.

In accordance with \cite{NN} there is a big variety of Hamiltonians
(\ref{B3}) ge\-ne\-rating non-equivalent continuous point symmetries of
equation~(\ref{B1}). The corresponding potential and mass terms are
defined up to arbitrary parameters or even up to arbitrary
functions.

In the present paper we consider the PDM systems defined up to
arbitrary parameters. Just such systems admit the most extended Lie
symmetries. Using the classification results presented in \cite{NZ} and~\cite{NN}
we enumerate these systems in the following Table~1, where $\varphi=\arctan\frac{x_2}{x_1}$ and the other Greek letters denote arbitrary constants parameters, which are supposed not to be zero simultaneously. Moreover, $\lambda$ and $\omega$ are either real or imaginary, the remaining parameters are real.

\vspace{1mm}

\begin{center} Table 1.
PDM systems with extended Lie symmetries.\end{center}

\vspace{1mm}

\begin{tabular}{ccll}
\hline
 No&$\begin{array}{c}\text{Inverse}  \ \text{mass} \ f
\end{array}$&Potential $V$&Symmetries\\
\hline

\vspace{2mm}

 1&$\big(r^2+1\big)^2
 $&\vspace{1mm}$-3 r^2$&$\begin{array}{l}M_{41}, M_{42}, M_{43},\\ M_{21}, M_{31}, M_{32}\end{array}$ \\
 2&\vspace{1mm}$\big(r^2-1\big)^2$&$-3 r^2$&$ \begin{array}{l}M_{01}, M_{02}, M_{03},\\ M_{21}, M_{31},
 M_{32}\end{array}$ \\
 3&\vspace{1mm}$x_3^2$&$\nu\ln(x_3)$&$\begin{array}{l}P_1,\ P_2,\ M_{12},\ D+\nu
t\end{array}$
\\
4&\vspace{1mm}$\tilde r^3$&$ \kappa x_3+\lambda\tilde r$&$
P_3+\kappa t,
\ D+\ri t\p_t, \ M_{12}$\\

5&\vspace{1mm}$x_1^3$&$\lambda x_1+\kappa x_3$&$P_3+\kappa t, \ P_2,
\
D+\ri t\p_t$\\
6&\vspace{1mm}$x_3^{\sigma+2}$&$\kappa
x_3^\sigma$&$\begin{array}{l}P_1,\ P_2,\ M_{12}, \ D+\ri\sigma t\p_t, \\
\sigma\neq
0,1,-2\end{array}$\\
7&\vspace{1mm}$\tilde r^{\sigma+2}{\rm e}^{\lambda\varphi}$&$\kappa \tilde
r^{\sigma}{\rm e}^{\lambda\varphi}$&$\begin{array}{l}M_{12}+\ri\lambda t\p_t,\
P_3,\\
D+\ri\sigma t\p_t, \ \sigma\neq0\end{array}$\\

 8& \vspace{1mm} $\tilde
r^2$&$
\frac{\lambda^2}2\varphi^2+\mu\varphi+\nu \ln(\tilde
r)
$& $ \begin{array}{l}B^1_1, B^1_2,\\D+\nu t, \ P_3 \end{array}
$\\
9&\vspace{1mm}$\tilde r^2{\rm e}^{\sigma \varphi}$&$\kappa
{\rm e}^{\sigma\varphi}+\frac{\omega^2}{2} {\rm e}^{-\sigma\varphi}$&$\begin{array}{l}N^1_1, \ N^1_2, P_3, \ D,\ K_3\end{array}$\\
10&\vspace{1mm}$ r^2$&$\nu\ln(r)+
{\frac{\lambda^2}2}\ln(r)^2$&$\begin{array}{l}B^2_1,\ B^2_2, \
 L_1, \ L_2, \ L_3 \end{array}$\\
 11&\vspace{1mm}$r^{2+\sigma}$&$\kappa r^{\sigma}+
\frac{\omega^2}{2}r^{-\sigma} $&$\begin{array}{l}N^2_1,\
N^2_2,\
 L_1, \ L_2, \ L_3 \end{array}$\\

 \hline\hline
 \end{tabular}

\vspace{2mm}

The symmetry operators presented in column 4 of the table are given by the following formulae:
\begin{gather}
 P_{i}=p_{i}=-i\frac{\partial}{\partial x_{i}},\quad D=x_n p_n-\tfrac{3\ri}2,\nonumber\\
M_{ij}=x_ip_j-x_jp_j,\quad M_{0i}=\tfrac12\big(K^i+P_i\big), \quad M_{4i}=\tfrac12\big(K^i+P_i\big),\nonumber \\
B^1_1=\lambda\sin (\lambda t) M_{12}
\big(\lambda^2\varphi+\nu\big)\cos (\lambda t), \quad B^1_2=\frac{\p}{\p t}B^1_1,\nonumber\\
 B^2_1= \sin(\lambda t)D- \cos(\lambda
t)\left({\lambda}\ln(r)+\tfrac\nu\lambda\right), \quad B^2_2=\frac{\p}{\p t}B^2_1,\nonumber\\
 N^1_1=\omega\cos
(\omega\sigma t) L_3-\sin(\omega\sigma t)\big(\ri
\p_t- \omega^2{\rm e}^{-\sigma\Theta}\big) ,\quad N^1_2=\frac{\p}{\p t}N^1_1,\nonumber \\
 N^2_1=\omega\cos(\omega\sigma
t)D+\sin(\omega\sigma t)\big(\ri \p_t-{\omega^2} r^{-\sigma}\big)
,\quad N^2_2=\frac{\p}{\p t}N^1_1,\label{SO}
\end{gather}
where $K_{i}=x_nx_n p_i -2x_iD$ and indices $i$, $j$, $k$, $n$ take the values $1$, $2$, $3$. In addition, all the presented systems admit symmetry operators $P_0=\ri\frac{\p}{ \p t}$ and the unit operator, the latter is requested to obtain the closed symmetry algebras.

Rather surprisingly, all systems
presented in Table 1 (except ones given in items 4 and 5 with $\kappa\neq0$) are exactly solvable. In the following sections we
present their exact solutions. To obtain these solutions we use some
nice properties of the considered systems like superintegrability and supersymmetry
with shape invariance. Let us remind that the quantum
mechanical system is called superintegrable if it admits more integrals
of motion than its number of degrees of freedom.
In accordance with Table 1 we can indicate 11 inequivalent PDM
systems which are defined up to arbitrary parameters and admit Lie
symmetry algebras of dimension five or higher. Notice that the systems
fixed in items 4 and 5 admit five dimension symmetry algebras while the
remaining systems admit more extended symmetries.
\section{ Systems with fixed mass and potentials.}
First we consider systems whose mass and potential terms are fixed, i.e., do not include arbitrary parameters.
 These systems are presented in items 1, 2 of Table~1 and others provided the mass does not depends on parameters and parameters of the potential are trivial.

\subsection{ System invariant w.r.t.\ algebra $\boldsymbol{\mathfrak{so}(4)}$.}
Consider Hamiltonian (\ref{B2}) with functions $f$ and $V$ presented
in item~1 of Table~1:
\begin{gather}\label{H2}H=\tfrac12p_a\big(1+r^2\big)^2p_a-3 r^2.\end{gather}
The eigenvalue problem for this Hamiltonian can be written in the
following form:
\begin{gather}\label{ep}H\psi= 2E\psi,\end{gather} where
$E$ are yet unknown eigenvalues.

Equation (\ref{ep}) admits six integrals of motion $M_{AB}$, $A,B=1,2,3,4$, presented in equation~(\ref{SO}). Let us write them explicitly
\begin{gather} M^{ab}=x^ap^b-x^bp^a,\quad
 M^{4a}=\tfrac12\big(r^2-1\big)p^a
-x^ax^bp^b+\tfrac{3\ri}2x^a.\label{im} \end{gather}

Operators (\ref{im}) form a
basis of algebra $\mathfrak{so}(4)$. Moreover, the first Casimir operator of this algebra is proportional to Hamiltonian (\ref{H2}) up to the constant shift
\begin{gather*}
C_1=\tfrac12M_{AB}M_{AB}=\tfrac12 ( H-9),\end{gather*}
while the second Casimir operator $C_2=\varepsilon_{ABCD}M_{AB}M_{CD}$ appears to be zero.

Thus like the Hydrogen atom system (\ref{ep}) admits six integrals of motion belonging to algebra $\mathfrak{so}(4)$ and is maximally superintegrable.

Using our knowledge of unitary representations of algebra $\mathfrak{so}(4)$ is possible to find eigenvalues~$E$ algebraically:
\begin{gather}\label{ev2} E=4 n^2+5,\end{gather}
where $n=0,1,2,\dots $ are natural numbers.

 To find the eigenvectors of Hamiltonian (\ref{H2}) corresponding to eigenvalues (\ref{ev2}) we use the rotation invariance of (\ref{ep}) and separate variables. Introducing spherical variables and expanding solutions via spherical functions
 \begin{gather}\label{rv}\psi=\frac1r\sum_{l,m}\phi_{lm}(r)Y^l_m\end{gather}
 we come to the following equations for radial functions
 \begin{gather*} \label{re}\left(-\big(r^2+1\big)^2\left(\frac{\p^2}{\p r^2}-\frac{l(l+1)}{r^2}\right)-4r\big(r^2+1\big)\frac{\p}{\p r}-2r^2\right)\varphi_{lm}\\
 \quad{} =\left(4 n^2+1\right)\varphi_{lm},
 \end{gather*} where $l=0,1,2,\dots$ are parameters numerating eigenvalues of the squared orbital momentum.
 The square integrable solutions of these equations are
 \begin{gather}\label{soll}\varphi_{lm}=C_{lm}^n\big(r^2+1\big)^{-n-\frac12}r^{l+1}
 {\cal F}\left([A,B],[C] -r^2\right),\end{gather}
 where
 \begin{gather*}A=-n+l+1, \quad B=-n+\tfrac12,\quad C= l+\tfrac32.\end{gather*}
 ${\cal F}(\cdots)$ is the hypergeometric function and $C_{lm}^n$ are integration constants. Solutions~(\ref{soll}) tend to zero at infinity provided $n$ is a natural number and $l\leq n-1$.

 Thus the system (\ref{ep}) is maximally superintegrable and exactly sol\-vable.

\subsection{ System invariant w.r.t.\ algebra $\boldsymbol{\mathfrak{so}(1,3)}$.}
 The next Hamiltonian we consider corresponds to functions~$f$ and~$V$
 presented in item~2 of Table~1.
The related eigenvalue problem includes the following equation
\begin{gather}\label{ep1} H\psi\equiv-\tfrac12\big(\p_a\big(1-r^2\big)^2\p_a+6 r^2\big)\psi=E\psi.\end{gather}

Equation (\ref{ep1}) admits six
integrals of motion $M_{\mu\nu}$, $\mu, \nu=0, 1,2,3,$ given by equation (\ref{SO}), which can be written explicitly in the following form
\begin{gather} M_{ab}=x^ap^b-x^bp^a, \nonumber\\
 M_{0a}=\tfrac12\big(r^2+1\big)p^a -x^ax^bp^b+\tfrac{3\ri}2x^a, \quad a,b=1,2,3.\label{im1} \end{gather}
These operators form a basis of algebra~$\mathfrak{so}(1,3)$, i.e., the Lie algebra of Lorentz group.

As in the previous section, the corresponding first Casimir operator is expressed via the Hamiltonian, namely
\begin{gather}\label{uh} C_1=\tfrac12M^{ab}M^{ab}-M^{0a}M^{0a}=\tfrac12 ( H+9),\end{gather} while the second one appears to be zero.

Using our knowledge of irreducible unitary representations of Lorentz group we find eigenvalues of $C_1$ and $C_2$ in the form
\cite{naimark,IM}:
\begin{gather*} c_1=1-j_0^2-j_1^2, \quad c_2=2\ri j_0j_1,\end{gather*} where $j_0$ and $j_1$ are quantum numbers labeling irreducible representations. Since the second Casimir operator $C_2$ is trivial, we have $c_1=j_0=0$. So there are two possibilities~\cite{naimark}: either~$j_1$ is an arbitrary imaginary number, and the corresponding representation belongs to the principal series, or $j_1$ is a real number satisfying $|j_1|\leq1 $, and we come to the subsidiary series of IRs. So
\begin{gather}\label{j1}j_1={\rm i}\lambda, \quad c_1=1-j_1^2=\lambda^2+1, \end{gather}
where $\lambda$ is an arbitrary real number, or, alternatively,
\begin{gather}\label{j2}0\leq j_1\leq 1, \quad c_1=1-j_1^2.\end{gather}

In accordance with (\ref{uh}) the related eigenvalues $ E$ in~(\ref{ep1}) are
\begin{gather}E=-5-j_1^2.\label{EE}\end{gather}

In view of the
rotational invariance of equation (\ref{ep1}) it is convenient to
represent solutions
in form~(\ref{rv}). As a result we obtain the following radial
equations
\begin{gather} \left(-\big(r^2-1\big)^2\left(\frac{\p^2}{\p
r^2}-\frac{l(l+1)}{r^2}\right)-4r\big(r^2-1\big)\frac{\p}{\p
r}-2r^2\right)\varphi_{lm}\nonumber\\
\quad {} =(\tilde E+4)\varphi_{lm}. \label{re1}\end{gather}

The general solution of (\ref{re1}) is
\begin{gather} \varphi_{lm}=\big(1-r^2\big)^{-
 \frac12-k}\big(C_{lm}^k r^{l+1}{\cal F}\big([A,B],[C],r^2\big) \nonumber \\
 \hphantom{\varphi_{lm}=}{} +\tilde C_{lm}^k r^{-l}{\cal F}\big([\tilde A,\tilde B],[\tilde C], r^2\big)\big),\label{soll1}
 \end{gather}
 where
 \begin{gather*} A=-k+l+1, \quad B= -k+\tfrac12,\quad C= l+\tfrac32,\\
  \tilde A = -k-l, \quad \tilde B=-k+\tfrac12,\quad \tilde C=\tfrac12-l, \quad k=\tfrac12\sqrt{-\tilde E-5}
  \end{gather*} and is singular at $r=1$. However, for $\tilde C_{lm}^k=0$ and $k= j_1$ the solutions are normalizable in some specific metric~\cite{NZ}.

 Thus the system presented in item 7 of Table~1 is exactly solvable too. The corresponding eigenvalues and eigenvectors are given by equations (\ref{j1}), (\ref{j2}), (\ref{EE}) and (\ref{soll1}) correspondingly.

\subsection{ Scale invariant systems.}
Consider one more PDM system which is presented in item~3 of the
table and includes the following Hamiltonian:
Let us note that the free fall effective potential appears also one more system specified in Table~1. Thus, considering the inverse mass and potential specified in item~3 we come to the following Hamiltonian
\begin{gather}\label{oj}\begin{split}&
H=-\frac12\left(x_3\frac{\p}{\p x_3}x_3\frac{\p}{\p x_3}+x_3\frac{\p}{\p x_3}+x_3^2\left(\frac{\p^2}{\p x_1^2}+\frac{\p^2}{\p x_2^2}\right)\right)\\&+\nu\ln(x_3)\end{split}.\end{gather}
Equation (\ref{ep1}) with Hamiltonian given in (\ref{oj}) can be easily solved by separation of variables in Cartesian coordinates. Expanding the wave function $\psi$ via eigenfunctions of integrals of motion $P_1$ and $P_2$:
\begin{gather}\label{buh}\psi=\exp(-\ri(k_1 x_1+k_2x_2))\Phi(k_1,k_2,x_3)\end{gather}
and introducing new variable $y=\ln(x_3)$ we come to the following equation for $\Phi=\Phi(k_1,k_2,x_3)$:
\begin{gather}\label{uaa}
-\frac{\p^2 \Phi}{\p{y^2 }}+\left(\big(k_1^2+k_2^2\big)\exp(2y)+2\nu y\right)\Phi=\tilde E\Phi \end{gather}
where $\tilde E=2E-\frac14$.

Here we consider the simplest version of equation (\ref{uaa}) when parameter $\nu$ is trivial:
 \begin{gather}\label{ua}
-\frac{\p^2 \Phi}{\p{y^2 }}+\big(k_1^2+k_2^2\big)\exp(2y)\Phi=\tilde E\Phi.\end{gather}
This equation is scale invariant and can be easily solved. Its square integrable solutions are given by Bessel functions
\begin{gather*}
\Psi=C^E_{k_1k_2}K_{\ri \sqrt{\tilde E}}\Big(\sqrt{k_1^2+k_2^2}\ln(x_3)\Big),\end{gather*}
where $C^E_{k_1k_2}$ are integration constants and $\tilde E$ are arbitrary real parameters.

It is interesting to note that there are rather non-trivial relations between the results given in the present and previous sections. Equation  (\ref{ua}) admits six integrals of motion which are nothing but the following operators{\samepage
\begin{gather}\label{onem}P_1,\ P_2,\ K_1,\ K_2, \ M_{12},\  D,\end{gather}
which are presented in equations (\ref{SO}).}

Like operators (\ref{im1}) integrals of motion (\ref{onem}) form a basis of the Lie algebra of Lorentz group, and we again can find the eigenvalues of Hamiltonian (\ref{ua}) algebraically by direct analogy with the above. We will not present this routine procedure since there exist strong equivalence relations between Hamiltonians (\ref{ua}) with zero $\nu$ and (\ref{H2}). To find them we note that basis (\ref{onem}) is equivalent to the following linear combinations of the basis elements:
\begin{gather} M_{01}, \ M_{02}, M_{04}, \ M_{41}\ M_{42}, \ M_{12},\label{mecem}\end{gather} whose expressions via operators~(\ref{onem}) are given by equation~(\ref{SO}). To reduce~(\ref{mecem}) to the set~(\ref{im1}) it is sufficient to change subindices 4 to 3, i.e., to make the rotation in the plane 43. The infinitesimal operator for such rotation is given by the following operator
\begin{gather*}M_{43}=\tfrac12(K_3+P_3)=\tfrac12\big(r^2-1\big)p_3
-x_3x_bp_b+\tfrac{3\ri}2x_3,
\end{gather*}
which belongs to the equivalence group of equations.
Solving the corresponding Lie equations and choosing the group parameter be equal $\frac{\pi}{2}$ we easily find the requested equivalence transformations.

One more scale invariant system is presented in item~8 where all parameters of potential are zero. The relation Hamiltonian looks as follows:
\begin{gather} \label{12} H=-\tilde r\frac{\p}{\p{x_\alpha}}\tilde r\frac{\p}{\p
{x_\alpha}} -x_\alpha\frac{\p}{\p{x_\alpha}}-\tilde r^2\frac{\p^2}{\p {x_3^2}},\quad \alpha=1,2.\end{gather}
                         Considering the eigenvalue problem for (\ref{12}) it is convenient to use the cylindrical variables
\begin{gather}\label{CV}\tilde r=\sqrt {x_1^2+x_2^2}, \ \varphi=\arctan\frac{x_2}{x_1},\ x_3=z\end{gather}
and expand solutions via eigenfunctions of $M^{12}$ and $P_3=-\ri\frac{\p}{\p z}$:
 \begin{gather}\label{CV1} \Psi=\exp[\ri(\kappa\varphi +\omega z)]\Phi_{\kappa\omega}(\tilde r),\quad \kappa=0,\pm1, \pm2,..., -\infty<\omega<\infty.\end{gather}
 As a result we come to the following equations for radial functions $\Phi=\Phi_{\kappa\omega}(\tilde r)$:
 \begin{gather*}-\left(\tilde r\frac{\p} {\p \tilde r}\tilde r\frac{\p}{\p \tilde r}+\tilde r\frac{\p} {\p \tilde r}+\omega^2\right)\Phi=(\tilde E-\kappa^2.)\Phi\end{gather*}
 Square integrable (with the weight $\tilde r$) solutions of this equation are:
 \begin{gather}\label{soso}\Phi_{\kappa\omega}=\frac1{\tilde r}J_\alpha(\omega \tilde r),\quad \alpha=\kappa^2+1-\tilde E \end{gather}
 where $J_\alpha(\omega \tilde r)$ is Bessel function of the first kind.
 Functions (\ref{soso}) are normalizable and disappear at $\tilde r=0$ provided $\alpha\leq0$. The rescaled energies  $\tilde E$ continuously take the values $\kappa^2\leq\tilde E\leq\infty$.

The last scale invariant system which we have to consider is fixed in item~10 where $\nu=\lambda=0$ We will do it later in Section

\section{ Systems defined up to arbitrary parameters.}
In previous section we present exact solutions for systems with
fixed potential and mass terms. In the following we deal with
the systems defined up to arbitrary parameters.

\subsection{ The system with oscillator effective potential.}
Let us consider equation (\ref{seq}) with $f$ and $V$ are
 functions fixed in item~10 of Table~1, i.e.,
\begin{gather*}
\ri\frac{\p \psi}{\p t}=
\left(-\frac12\frac{\p}{\p{x_a}} r^{2}\frac{\p}{\p{x_a}}+\nu\ln(r)+
{\frac{\lambda^2}2}\ln(r)^2\right)\psi.\end{gather*}

These equations admit extended symmetries Lie symmetries (whose generators are indicated in the table) being
invariant w.r.t.\ six-pa\-ra\-met\-ri\-cal Lie group. Let us show that they also admit hidden supersymmetries.

In view of the rotational invariance and
symmetry of the considered equations with respect to shifts of time variable, it is reasonable to
to search for their solutions in spherical variables, i.e., in the
following form
\begin{gather}\label{psi}\Psi={\rm e}^{-iEt}R_{lm}( r)Y_{lm}(\varphi,\theta),\end{gather}
where $\varphi$ and $\theta$ are angular variables and
$Y_{lm}(\varphi,\varphi)$ are spherical functions, i.e.,
eigenvectors of $L^2=L_1^2+L_2^2+M_{12}^2$ and $ M_{12}$. As a result we
come to the following radial equations
\begin{gather} \left(-r\frac{\p R_{lm}}{\p r}r\frac{\p R_{lm}}{\p r}-r\frac{\p R_{lm}}{\p r}\right.\nonumber\\\left.
\quad {} +{l(l+1)}+\nu\ln(r)+ {\frac{\lambda^2}2}\ln(r)^2\right)R_{lm}
=2ER_{lm}.\label{equq7} \end{gather}

Introducing new variable $y=\sqrt{2}\ln (r)$ we can rewrite equation (\ref{equq7}) in the following form:
\begin{gather}\label{ree}\left(-\frac{\p^2}{\p y^2}+l(l+1)+\nu y+\frac{\lambda^2}2y^2\right)R_{lm}(y)=\tilde ER_{lm}(y),\end{gather}
where $\tilde E=E-\frac14$.

Let $\lambda\neq0$ then equation (\ref{ree}) is reduced to the 1D harmonic oscillator up to the additional term $l(l+1)$ The admissible eigenvalues $\tilde E$ are given by the following formula
\begin{gather*}\tilde E =n+l(l+1).\end{gather*}
where $n$ is a natural number.
The corresponding eigenfunctions are well known and we will not presented them here. The same is true for supersymmetric aspects of the considered system.

If parameter $\lambda$ is equal to zero then (\ref{ree}) reduces to equation with free fall potential slightly modified by the term $l(l+1)$. The corresponding solutions can be found in textbooks devoted to quantum mechanics.

\subsection{ The systems with potentials equivalent to 3d oscillator.}
Consider now the system represented in item~11 of the table. The corresponding equation (\ref{seq}) takes the following form:
\begin{gather}\label{equ5}\ri\frac{\p \psi}{\p t}=
\left(-\frac12\p_a r^{\sigma+2}\p_a+\kappa r^{2\sigma}+
\frac{\omega^2}{r^{2\sigma}}\right)\psi.\end{gather}
Like in previous section we represent the wave function in the form given in (\ref{psi}) and came to the following radial equation
\begin{gather} -r^{2\sigma+2}\frac{\p^2R_{lm}}{\p r^2}-(2\sigma+4)r^{2\sigma+1}\frac{\p R_{lm}}{\p r}\nonumber\\
\quad {}  +
\big(r^{2\sigma}({l(l+1)+\kappa})+\omega^2r^{-2\sigma}\big)R_{lm}
=2ER_{lm}.\label{equ7} \end{gather}

Using the Liouville transform
\begin{gather*}
r\to z=r^{-\sigma},\quad R_{lm}\to\tilde R_{lm}=z^{\frac{\sigma+3}{2\sigma}}R_{lm},\end{gather*} we reduce (\ref{equ7}) to the following form
 \begin{gather}\label{equ9}-\sigma^2\frac{\p^2\tilde R_{lm}}{\p z^2}+\left(\frac{l(l+1)+\delta}{z^2}+\omega^2z^2\right)\tilde R_{lm}
=2E\tilde R_{lm},\end{gather} where
$ \delta=\frac3{4}(\sigma+1)(\sigma+3)+
2\kappa$.

Equation (\ref{equ9}) describes a deformed 3d harmonic oscillator
including two deformation parameters, namely, $\sigma$ and $\kappa$.

Let
\begin{gather}\label{con}2\kappa=-\sigma^2-3\sigma-2,\end{gather}
then equation (\ref{equ9}) is reduced to the following form
 \begin{gather}\label{eg1}H_l\tilde R_{lm}\equiv\left(-\sigma^2\frac{\p^2}{\p z^2}+\frac{(2l+1)^2
 {-\sigma^2}}{4z^2}+\omega^2z^2\right)\tilde R_{lm}
=2E\tilde R_{lm}.\end{gather}

Equation (\ref{eg1}) is shape invariant. Hamiltonian~$H_r$ can be factorized
\begin{gather}H_l=a_l^+a_l-C_l,\label{eg2}\end{gather}
where
\begin{gather*} a=-\sigma\frac{\p}{\p z}+W,\quad a^+=\sigma\frac{\p}{\p z}+W,\nonumber\\
W=\frac{2l+1+\sigma}{2z}+\omega z,\quad C_l=\omega(2l+2\sigma+1).
\end{gather*}

The superpartner $\hat H_l $ of Hamiltonian~(\ref{eg2}) has the following property
\begin{gather*}\hat H_l\equiv a_la_l^++C_l=H_{l+\sigma}+C_l.
\end{gather*}
Thus our Hamiltonian is shape invariant.

Thus to solve equation (\ref{eg1}) we can use the standard tools of SUSY quantum mechanics and find the admissible eigenvalues in the following form{\samepage
 \begin{gather}\label{eg6}E_n=\omega\left(2n\sigma+l+
 \sigma+\tfrac12\right)=\omega\left(2n+l+\tfrac32\right)+\delta\omega(2n+1),\end{gather}
 where $\delta=\sigma-1$.}

 Equation (\ref{eg6}) represents the spectrum of 3d isotropic harmonic oscillator deformed by the term proportional to $\delta$.

For equation (\ref{equ9}) we obtain in the analogous way
\begin{gather}\label{spect}E_n=\frac\omega2\big(\sigma(2n+1)+
\sqrt{(2l+1)^2+\tilde \kappa}\big),\end{gather} where $\tilde
\kappa=8(\kappa+1)+\sigma(\sigma+3).$
The related eigenvectors are
expressed via the confluent hypergeometric functions $\cal F$:
\begin{gather*}
R_n={\rm e}^{-\frac{\omega r^{\sigma}}{2\sigma}}r^{\sigma n-\frac{E_n}{\omega}}{\cal F}\left(-n, \frac{E_n}{\sigma\omega}-n,\frac{\omega}{\sigma} r^{-\sigma}\right),
\end{gather*}
where $n$ is integer and $E_n$ is eigenvalue (\ref{spect}).

\subsection{ System with angular oscillator potential.}
The next system which we consider is specified by the inverse mass and potential presented in item~8 of the table. The corresponding Hamiltonian is:
\begin{gather*}
H=p_a r^2p_a +\frac{\lambda^2}2\varphi^2+\sigma\varphi+\nu \ln(\tilde r).\end{gather*}
The corresponding eigenvalue equation is separable in cylindrical variables, thus it is reasonable to represent the wave function as follows
\begin{gather}\label{cv} \psi=\Psi(\tilde r)\Phi(\varphi)\exp(-{\rm i}kx_3).\end{gather}
As a result we obtain the following equations for radial and angular variables
\begin{gather}\label{ep6}\left(-\tilde r\p_{\tilde r}\tilde r\p_{\tilde r}-\tilde r\p_{\tilde r}+\nu\ln(\tilde r)+k^2\tilde r^2-\mu\right)\Psi(\tilde r)=0\end{gather}
and
\begin{gather}\label{ep7}\left(-\frac{\p^2}{\p \varphi^2}+\frac{\lambda^2}{2}\varphi^2+\sigma\varphi-
\mu\right)\Phi(\varphi)=0,
\end{gather}
where $\mu$ is a separation constant.

For $\lambda$ nonzero equation (\ref{ep7}) is equivalent to the Harmonic oscillator. The specificity of this system is that, in contrast with (\ref{ree}), it includes angular variable $\varphi$ whose origin is \begin{gather}\label{uhuhuh}0\leq\varphi\leq2\pi.\end{gather}

For trivial $\lambda$ our equation (\ref{ep7}) is reduced to equation with free fall potential, but again for the angular variable satisfying (\ref{uhuhuh}).

The radial equation (\ref{ep6}) is simple solvable too. In the case $k=0$ we again come to the free fall potential.

\subsection{ Systems with Morse  effective potential.}
The next system we consider is specified by the inverse mass and potentials represented in item~9 of Table~1. The corresponding Hamiltonian is
\begin{gather*}
H= -\frac{\p}{\p x_a}\tilde r^2{\rm e}^{\sigma \varphi}\frac{\p}{\p x_a}+\kappa
{\rm e}^{\sigma\varphi}+\frac{\omega^2}{2} {\rm e}^{-\sigma\varphi}.\end{gather*}

Introducing again the cylindric variables and representing the wave function in the form (\ref{cv}) we come to the following equations for the radial and angular variables
\begin{gather*}
\left(-\left(\frac{\p^2}{ \p y^2}+\frac{\p}{ \p y}\right)+\mu+k^2{\rm e}^{2y}\right)\Psi(\tilde r)=\mu\Psi(\tilde r)\end{gather*}
and
\begin{gather}\label{olala}
\left(-{\rm e}^{\sigma\varphi}\left(\frac{\p^2}{ \p \varphi^2}+\kappa-\mu\right)
+\frac{\omega^2}{2} {\rm e}^{-\sigma\varphi}\right)\Phi(\varphi)=\tilde E\Phi(\varphi ).\end{gather}

Dividing all terms in (\ref{olala}) by $\exp({\sigma\varphi})$ we obtain the following equation:
\begin{gather*}
\left(-\left(\frac{\p^2}{ \p \varphi^2}+\kappa-\mu\right)
+\frac{\omega^2}{2} {\rm e}^{-2\sigma\varphi} \right)\Phi(\varphi)={\rm e}^{-\sigma\varphi} \tilde E\Phi(\varphi ).\end{gather*}
or
\begin{gather}\label{olalala}
\left(-\left(\frac{\p^2}{ \p \varphi^2}\right)
+\frac{\omega^2}{2} {\rm e}^{-2\sigma\varphi}-\tilde E{\rm e}^{-\sigma\varphi}\right)\Phi(\varphi )=\hat E\Phi(\varphi ),\end{gather}
where we denote $\hat E=\mu-\kappa$.

Formula (\ref{olalala}) represents the Schr\"odinger equation with Morse potential. This equation is shape invariant and also can be solved using tools of SUSY quantum mechanics. We demonstrate this procedure using another system.

Considering the mass and potential presented in item~6 of Table 1 we come to the following Hamiltonian
\begin{gather*}H=\frac12p_ax_3^{\sigma+2}p_a+\kappa x_3^\sigma.\end{gather*}
Equation (\ref{ep1}) with Hamiltonian  (\ref{oj}) can be solved by separation of variables in Cartesian coordinates. Expanding the wave function $\psi$ via eigenfunctions of integrals of motion $P_1$ and $P_2$ in the form (\ref{buh}) and introducing new variable $y=\ln(x_3)$
 we reduce the problem   to the following equation for $\Phi(k_1,k_2,x_3)$:
\begin{gather}\label{lll}\left(-\frac{\p}{\p{x_3}}x_3^{\sigma+2}\frac{\p}{\p{x_3}}+
x_3^{\sigma+2}k^2+2\kappa x_3^\sigma\right)\Phi=2E\Phi\end{gather}
were $k^2=k_1^2+k_2^2$.

Dividing all terms in (\ref{lll}) by $x_3^\sigma$ we can rewrite it in the following form:
\begin{gather*}\left(-\frac{\p^2}{\p y^2}-(\sigma+1)\frac{\p}{\p y}-2E\exp(-\sigma y)+k^2\exp(2 y)+2\kappa\right)\Phi=0 \end{gather*}

In the particular case $\sigma=2$ we again come to the equation with Morse effective potential.

One more system which can be related to Morse potential is represented in item~7 and include the following Hamiltonian:
\begin{gather*}H=\frac12p_a\exp(\lambda\varphi)\tilde r^{\sigma+2}p_a+\nu\exp(\lambda\varphi)\tilde r^{\sigma}.\end{gather*}
The corresponding equation (\ref{ep1}) is separable in the cylindrical variables (\ref{CV}) provided $\sigma\cdot\lambda=0$ and again includes the Morse effective potential.

Let us return to equation (\ref{equ7}) and solve it using approach analogous to the presented above. In other words, we will change the roles of eigenvalues and coupling constants.

First we divide all terms in by $r^{2\sigma}$
and obtain
\begin{gather}-r^{2}\frac{\p^2R_{lm}}{\p r^2}-(2\sigma+4)r\frac{\p R_{lm}}{\p r} \nonumber\\
\quad {} +
\big(\omega^2r^{-4\sigma} +\mu r^{-2\sigma}\big)R_{lm}=\varepsilon R_{lm},\label{equq1}\end{gather} where
\begin{gather}
\label{B}\varepsilon={-l(l+1)-2\kappa}, \quad \mu=-2E.\end{gather}

 Applying the Liouville transform
\begin{gather*}
r\to \rho=\ln(r),\quad R_{lm}\to\tilde R_{lm}={\rm e}^{-\frac{\sigma+3}2}R_{lm}\end{gather*}
we reduce (\ref{equq1}) to a more compact form
\begin{gather}\label{equq3} H_\nu\tilde R_{lm}\equiv\left(\!-\frac{\p^2}{\p \rho^2}+\omega^2{\rm e}^{-2\sigma\rho} +
(2\omega\nu+\omega{\sigma}){\rm e}^{-\sigma\rho}\!\right) \tilde
R_{lm}=\hat\varepsilon\tilde R_{lm},\!\!\! \end{gather} where
\begin{gather}\label{A}\hat\varepsilon=\varepsilon-
\left(\frac{\sigma+3}2\right)^2 ,\quad
\nu=\frac{\mu}{2\omega}-\frac{\sigma}2.
\end{gather}

Like (\ref{olalala}) equation (\ref{equq3}) includes the familiar Morse potential and so
is shape invariant. Indeed, denoting $\mu=2\omega(\nu+\frac\sigma2)$
we can factorize hamiltonian $H_\nu$ like it was done in (\ref{eg2})
where index $l$ should be changed to $\nu$ and{\samepage
\begin{gather*}\label{w}
W=\nu-\omega{\rm e}^{-a\rho},\quad C_\nu=\nu^2
\end{gather*}
and the shape invariance is easy recognized.}

To find the admissible eigenvalues $\varepsilon$ and the
corresponding eigenvectors we can directly use the results presented
in paper~\cite{Khare}, see item~4 of Table~4.1 there
\begin{gather*}
 \hat\varepsilon =\hat\varepsilon_n=-(\nu-n\sigma)^2,\quad
\big(\tilde R_{lm}\big)_n=y^{\frac\nu\sigma-
n}{\rm e}^{-\frac{y}2}L_n^{2(\frac\nu\sigma-n)}
(y), \end{gather*} where $y=\frac{2\omega}\sigma
r^{-\sigma}$.

Thus we find the admissible values of $\hat\varepsilon_n$. Using
definitions (\ref{B}) and~(\ref{A}) we can find the corresponding
values of $E$ which are in perfect accordance with~(\ref{spect}).

\section{ Discussion.}
The results presented above in Section 2 include the
complete list of continuous symmetries which can be admitted by PDM
Schr\"odinger equations, provided these equations are defined up to arbitrary
parameters.

It is important to note that the list of symmetries presented in the
fourth column of the table is valid only for the case of nonzero parameters
defining the potential and mass terms. If some (or all) of these parameters
are trivial, the corresponding PDM Schr\"odinger equation can have more
extended set of symmetries. For example, it is the case for the potential
and PDM presented in item 3 of the table, compare the list of symmetries
presented in column 4 with (24). The completed list of non-equivalent
symmetries can be found in [13] which generalizes the Boyer results [3] to
the case of PDM Schr\"odinger equations. As other extensions of results
of [3] we can mention the group classification of the nonlinear Schr\"odinger
equations [15] and the analysis of its conditional symmetries [6].

Thanks to their extended symmetries the majority of the presented
systems is exactly solvable. In Sections 3 and 4 we present the corresponding
solutions explicitly and discuss supersymmetric aspects of
some of them. However, two of the presented systems (whose mass and
potential are presented in items 4 and 5 of Table 1) are not separable, if
 arbitrary parameter $\kappa$ is nonzero. And just these systems
have "small" symmetry, admitting five parametrical Lie groups. For $\kappa$ equal to zero these systems are reduced to particular cases presented in items 6 and 11.

On the other hand, all systems admitting
six- or higher-dimensional Lie symmetry algebras are separable and
exactly solvable.
In addition to the symmetry under the six parameter Lie group, equation
(32) (which we call deformed 3d isotropic harmonic oscillator) possesses
a hidden dynamical symmetry w.r.t. group SO(1, 2). The effective
radial Hamiltonian is shape invariant, and its eigenvalues can be found
algebraically. In spite on the qualitative difference of its spectra (37)
and (38) of the standard 3d oscillator, it keeps the main supersymmetric
properties of the latter. We note that the shape invariance of PDM
problems usually attends their extended symmetries.

\end{document}